\documentclass[twocolumn,pre,floats,aps,amsmath,amssymb,showkeys]{revtex4}
\usepackage{graphicx}
\usepackage{epstopdf}
\usepackage{color}
\usepackage{times}
\usepackage{soul}
\DeclareGraphicsExtensions{.pdf,.eps,.png,.jpg,.mps}

\begin{document}


\title{Nonlinear edge modes in a honeycomb electrical lattice near the Dirac points}

\author{F. Palmero $^{1}$, L. Q. English $^{2}$,  J. Cuevas-Maraver $^{3}$, and P.G. Kevrekidis $^{4,5}$}

\affiliation{$^{1}$Grupo de F\'{\i}sica No Lineal, Departamento de F\'{\i}sica Aplicada I,
Escuela T\'{e}cnica Superior de Ingenier\'{\i}a Inform\'{a}tica, Universidad
de Sevilla, Avda Reina Mercedes s/n, E-41012 Sevilla, Spain}

\affiliation{$^{2}$Department of Physics and Astronomy
Dickinson College, Carlisle, Pennsylvania, 17013, USA}

\affiliation{$^{3}$Grupo de F\'{\i}sica No Lineal, Departamento de F\'{\i}sica Aplicada I,
Universidad de Sevilla. Escuela Polit\'{e}cnica Superior, C/ Virgen de Africa, 7, 41011-Sevilla, Spain.\\
Instituto de Matem\'{a}ticas de la Universidad de Sevilla (IMUS). Edificio
Celestino Mutis. Avda. Reina Mercedes s/n, 41012-Sevilla, Spain, Avda Reina Mercedes s/n, E-41012 Sevilla, Spain}

\affiliation{$^{4}$ Department of Mathematics and Statistics, University of Massachusetts,
Amherst MA 01003-4515, USA}
\affiliation{$^{5}$ Mathematical Institute, University of Oxford, Oxford, OX2
  6GG, UK}

\date{\today}

\begin{abstract}
  We examine - both experimentally and numerically - a two-dimensional
  nonlinear driven electrical lattice with honeycomb structure.
Drives are considered over a range of frequencies both outside (below and
above) and inside the band of linear modes.
  We identify a number of discrete breathers both existing in the bulk and also
  (predominantly) ones arising at the domain boundaries,
  localized either along the arm-chair or along the zig-zag edges.
   The types of edge-localized breathers observed and computed
  emerge in distinct frequency bands near the Dirac-point frequency of
  the dispersion
  surface while driving the lattice subharmonically (in a spatially
  homogeneous manner).
  These observations/computations can represent a starting point towards the
  exploration of the interplay of nonlinearity and topology in an
  experimentally tractable system such as the honeycomb electrical lattice.
\end{abstract}

\maketitle

\section{Introduction}

Honeycomb lattices have attracted substantial interest within the
physics community in recent years, due to their inherent potential of
topological surface phenomena \cite{bernevig}.
The interplay of topology and wave dynamics (both at the linear and
more recently at the nonlinear level) has had significant impact both
in the realm of optics~\cite{joann} and in that of acoustic/mechanical
systems~\cite{suss,ma}. Nonlinearity further adds to the complexity
and the wealth of this interplay, especially since in settings
such as optics~\cite{HuangNature,Peleg,Ablowitz1,Ablowitz2}
and atomic physics~\cite{carr1,carr2}, it emerges spontaneously
for large amplitude/density  excitations.

In two dimensions (2D) the foremost example of honeycomb material is,
of course, graphene~\cite{grap1,grap2,grap3}. One of the most
intensely studied macroscopic analogues is photonic graphene, and
edge-localized states were soon predicted to exist in such materials
\cite{savin,savin2,ablowitz,koh}. Furthermore, the role of
nonlinearity in this context is beginning to be examined (see
e.g.,~\cite{savin,savin2}), yet there are still numerous avenues worth exploring in this context relating to
the impact of nonlinearity, especially in experimentally
tractable settings.
In the linear regime, some pioneering experimental results have appeared in the literature
in photonic graphene that show the existence of edge-localized states
\cite{plotnik, noh}, and these were even shown to propagate in one
direction only, upon breaking the time-reversal symmetry
\cite{mikael}. Yet, we argue that the identification of unprecedented,
experimentally controlled settings where nonlinear states (both bulk
and edge ones) can be obtained is of value to the efforts to
understand nonlinear topological structures and how they differ from
their more standard, non-topological variants (as well as how such
states vary from linear topological ones). In that vein, we propose
as a platform worth exploring the setting of honeycomb electrical lattices.

More concretely, in this paper we report on a series of findings of
nonlinear
waves in a 2D electrical honeycomb lattice. That
intrinsic localized modes, also known as {\em discrete breathers} (DBs) can exist in square lattices of this kind has
been shown previously \cite{electric1}.
In fact, such modes are well-known to exist in a wide range of
physical
settings, summarized, e.g., in a number of reviews~\cite{Flach,Aubry}.
Here, however, we focus on the role
of the honeycomb geometry and drive the system over a wide
range of frequencies both within as well as outside the band of
small amplitude excitations. We find both experimentally and numerically
that not only can bulk-localized modes be identified in this setting, but also
edge-localized modes can be excited with a spatially homogenous,
subharmonic driver. These DBs bear frequencies around that of the
Dirac points. The exact DB frequency depends on the wave amplitude,
as expected from a soft nonlinear system, but interestingly also on
the type of edge. The relevant zig-zag edge-localized DBs are found to exist within
a frequency band that is
higher, in terms of frequencies (and non-overlapping) compared to the
arm-chair mode band. We
complement these results with a numerical stability analysis and find that these
discrete breathers do not appear to derive from a continuation of
linear modes, but that they come into existence via (saddle-node)
bifurcation phenomena. Our findings constitute a first step towards
the more systematic examination of stable bulk and edge modes
in such honeycomb electrical lattices and we hope will spurt further
efforts in this direction.

Our presentation is structured as follows. In section II we present
the mathematical model associated with the experimental setting of
interest, i.e., the honeycomb lattice of LC resonators. The underlying
linear modes are identified and their band is obtained for parameters
associated within the experimental range in Section III. Subsequently, in section
IV, we present an anthology of experimental and numerical results for similar
conditions between the experiment and the numerical computation.
The findings are presented for different values of the
driver frequency,
progressively moving from frequencies below the band to
ones above the band of linear states. Finally, we summarize our
findings and present our conclusions and some challenges arising
towards future work in Section V.

\section{The Model}

The experimental system investigated in this paper is a honeycomb lattice consisting of unit cells that are comprised of
LC resonators, whose nonlinearity is originated by using a varactor diode instead of the standard capacitor. These nonlinear resonators are coupled together into a two-dimensional lattice via coupling inductors. Such a system was studied in a previous publication \cite{electric1}, where it was found that stable two-dimensional ILMs/discrete breathers could be produced. That study used periodic boundary conditions exclusively, thus eliminating any lattice edges. In the present study, we have used free-ends boundary conditions, allowing a pair of both zig-zag and armchair edges. This has permitted us to investigate the dynamical interplay between lattice edges and nonlinear localized states.

More concretely, our use of a varactor diode (NTE 618) introduces a
specific (experimentally determined) nonlinear capacitance $C(V)$. We
also use inductors of value $L_2 = 330 \mu$H, and the resulting unit
cells are driven by a {periodic voltage source $\mathcal{E}(t)$ of frequency $f$}
{(i.e., the driving is {\em uniform})} via a resistor $R = 10 $k$\Omega$. Each single unit is coupled to its three neighbors via inductors $L_1=680 \mu$H building a honeycomb lattice.

Using basic circuit theory, the system can be described by the equations \cite{electric1, electric2},

\begin{eqnarray}
\frac{d i_{n,m}}{d \tau} & =& \frac{L_2}{L_1} \left( \sum_{j,k} v_{j,k}-K_{n,m}v_{n,m} \right)-v_{n,m} \label{lattice_eq} \\
\frac{d v_{n,m}}{d \tau} & =& \frac{1}{c(v_{n,m})}\left[ i_{n,m}-i^D(v_{n,m})-\right. \nonumber \\
& & \left. \frac{v_{n,m}}{C_0\omega_0R_e}+ {\frac{1}{ C_0\omega_0 R} \frac{\mathcal{E}(\Omega\tau)}{V_d}} \right], \nonumber
\end{eqnarray}
where the sum $(j,k)$ is taken over all neighbors of the $(n,m)$ node and $K_{n,m}$ is the number of
neighbors of node $(n,m)$. $K_{n,m}$ is equal to three in an infinite
lattice (or finite lattice with periodic boundary conditions), but in
a finite lattice with free boundaries it could be either $K_{n,m}=1$
or $K_{n,m}=2$ on the edges, depending on the particular lattice
node. {The varactor can be modeled as a nonlinear resistance in parallel with a nonlinear capacitance. As shown in \cite{electric2}, the nonlinear current $I^D(V)$ is given by
\begin{equation}	
I^D(V)=-I_s \exp(-\beta V) ,
\end{equation}
where $\beta=38.8$ V$^{-1}$ and $I_s=1.25 \times 10^{-14}$ A, and its capacitance $C(V)$ as
\begin{equation}
C(V) =
\begin{cases}
C_v+C_1(V')+C_2 (V')^2 & \text{if } V \leq V_c, \\
C_0 e^{-\alpha V}       & \text{if } V > V_c,
\end{cases}
\end{equation}
where $V'=(V-V_c)$, $C_0=788$ pF, $\alpha=0.456$ V$^{-1}$, $C_v=C_0 \exp(-\alpha V_c)$, $C_1=-\alpha C_v$, $C_2=100$ nF and $V_c=-0.28$ V.}

{The following dimensionless variables were used in
Eq.~(\ref{lattice_eq}): $\tau=\omega_0 t$, where
$\omega_0=1/\sqrt{L_2C_0}$; $\Omega=2\pi f/\omega_0$ is the dimensionless driving frequency; the dimensionless voltage
{$v_{n,m}=V_{n,m}/V_d$}, with $V_d$ representing the voltage amplitude of the driving; $i_{n,m}=(I_v-I_2)/(C_0\omega_0 V_d)$, where $I_v$ is the full current through the unit cell and $I_2$ the current through the inductor $L_2$, both corresponding to cell $(n,m)$ and $i^D=I^D/(C_0 \omega_0 V_d)$. A phenomenological dissipation resistor, $R_l$, was included in the model to better approximate the experimental dynamics and $R_e$ is the equivalent resistance so $1/R_e=1/R+1/R_l$. In all cases, the ratio $L_2/L_1$ characterizes the strength of the effective discreteness of the system (with the uncoupled limit obtained for $L_1 \rightarrow \infty$). We should add that this is still only a simplified model of the varactor diodes, and comparison between theoretical and experimental results will not be exact. Yet, it is an important first step in the modeling effort towards understanding this setup.}

\section{Linear modes}

\begin{figure}
\includegraphics[width=0.4\textwidth]{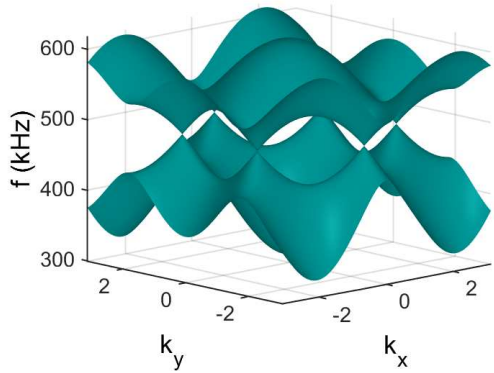}
\caption{Typical band structure of the infinite honeycomb lattice in the first Brioullin zone. There are six Dirac points corresponding to
$(k_x,k_y)=(0,\pm 4 \pi/(3\sqrt{3})),(2\pi/3,\pm 2\pi/(3\sqrt{3}))$,  $(-2\pi/3,\pm 2\pi/(3\sqrt{3}))$ and a
frequency $f_d =  489.11$ kHz, cf. Eq. (\ref{eq:disprel}).}
\label{band}
\end{figure}

In the linear limit ($c(v)=1$, $i_d=0$) the undriven and undamped system reduces to
\begin{equation}
 \frac{d ^2 v_{n,m}}{d \tau^2}=\frac{L_2}{L_1}\left( \sum_{j,k} v_{j,k}-K_{n,m}v_{n,m} \right)-v_{n,m}.
 \label{lineal}
\end{equation}
Linear modes can be found as plane-wave solutions. An infinite lattice (with nearest neighbor spacing of $1$) can be generated from lattice vectors $\mathbf{e}_\pm=(1/2,\pm\sqrt{3}/2)$ (see e.g. \cite{Cserti}), and the dispersion relation $\omega(\mathbf{k})$, with $\mathbf{k}=(k_x,k_y)$, is given by
\begin{eqnarray}\label{eq:disprel}
 \omega^2 & = & \frac{1}{C_0 L_2}+\frac{1}{C_0 L_1}\left[ 3 \pm \right.  \\
  & & \left. \sqrt{3+2 \cos(\sqrt{3} k_y)+4\cos(3k_x/2)\cos(\sqrt3k_y/2)}\right] \nonumber,
\end{eqnarray}
which yields a band of frequencies between $f_\mathrm{min}=\sqrt{1/(C_0L_2)}/(2\pi)\approx 312$ kHz and $f_\mathrm{max}=\sqrt{1/(C_0L_2)+6/(C_0L_1)}/(2\pi)\approx $ 617 kHz. As shown in Fig.~\ref{band}, the band structure corresponds to a graphene-like surface where six Dirac points exist at a frequency of $\omega_d=\sqrt{3/(C_0L_1)+1/(C_0L_2)}$, or
$f_d=\omega_d/2\pi \approx 489.11$ kHz.

In a finite lattice, wave vectors $\mathbf{k}$ are quantized. However, this quantization depends on the boundary conditions and the way the lattice is tiled. Because of this, one must be very cautious with the choice of boundary conditions, the way the honeycomb is generated and the lattice size if the Dirac point is intended to be in the linear mode spectrum. For periodic boundary conditions, an explicit expression of the eigenfrequencies can be attained \cite{Cserti}, but, for free ends boundary conditions, one must rely on the numerical solution of Eq.~(\ref{lineal}) for getting the linear mode spectrum.

In the present study, experimental limitations restrict us to a lattice of $6\times6$ nodes, distributed as shown in Fig.~\ref{lattice}. The boundaries are free, as we are interested in seeking  edge-localized breathers, as shown below. With this particular choice, there is a sole eigenmode oscillating with the Dirac frequency $f=f_d=489.11$ Hz. Figure \ref{lattice} also shows the oscillation pattern of such eigenmode (Dirac mode), which is similar to the oscillation pattern in an infinite lattice. {We have checked that this sole Dirac mode is present when tiling this lattice to a larger one with $6N\times6M$ with $(N,M)\in\mathbb{N}$ nodes}.

In addition, the structure of the edges of our lattice can be crucial for the formation of edge-localized breathers. According to the types of edge modes on graphene-like systems \cite{koh}, our system should be able to support both vertical zigzag and horizontal armchair edge-localized breathers.

\begin{figure}
\includegraphics[width=0.4\textwidth]{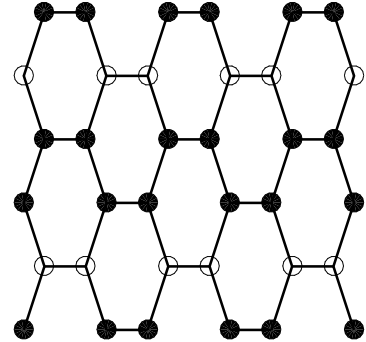}
\caption{Finite size ($6 \times 6$) lattice with free boundaries; shown is the Dirac mode, i.e. the linear mode corresponding to $f=f_D=489.11$ kHz. Black and white circles correspond, respectively, to a normalized amplitude in $t=0$ of $1/2$ and $-1$.}
\label{lattice}
\end{figure}

\section{Nonlinear modes: numerical and experimental results}

In this section, we will describe some numerical and experimental
  results on the existence of DBs when the electric lattice is driven
  uniformly. We have observed two kinds of such DBs, depending on
  whether they are localized on the lattice boundaries or elsewhere.
  We will call these DBs edge breathers (EBs) or bulk breathers (BBs),
  respectively, hereafter. The latter owe their existence to the intrinsic
  nonlinearity of the lattice (see e.g. \cite{electric1,electric2}).
  In the former case, there is an interplay between nonlinearity and
  the nature of the coupling in the vicinity of the boundary.

\subsection{Driving near the lowest frequency mode}

\begin{figure}
\includegraphics[width=0.23\textwidth]{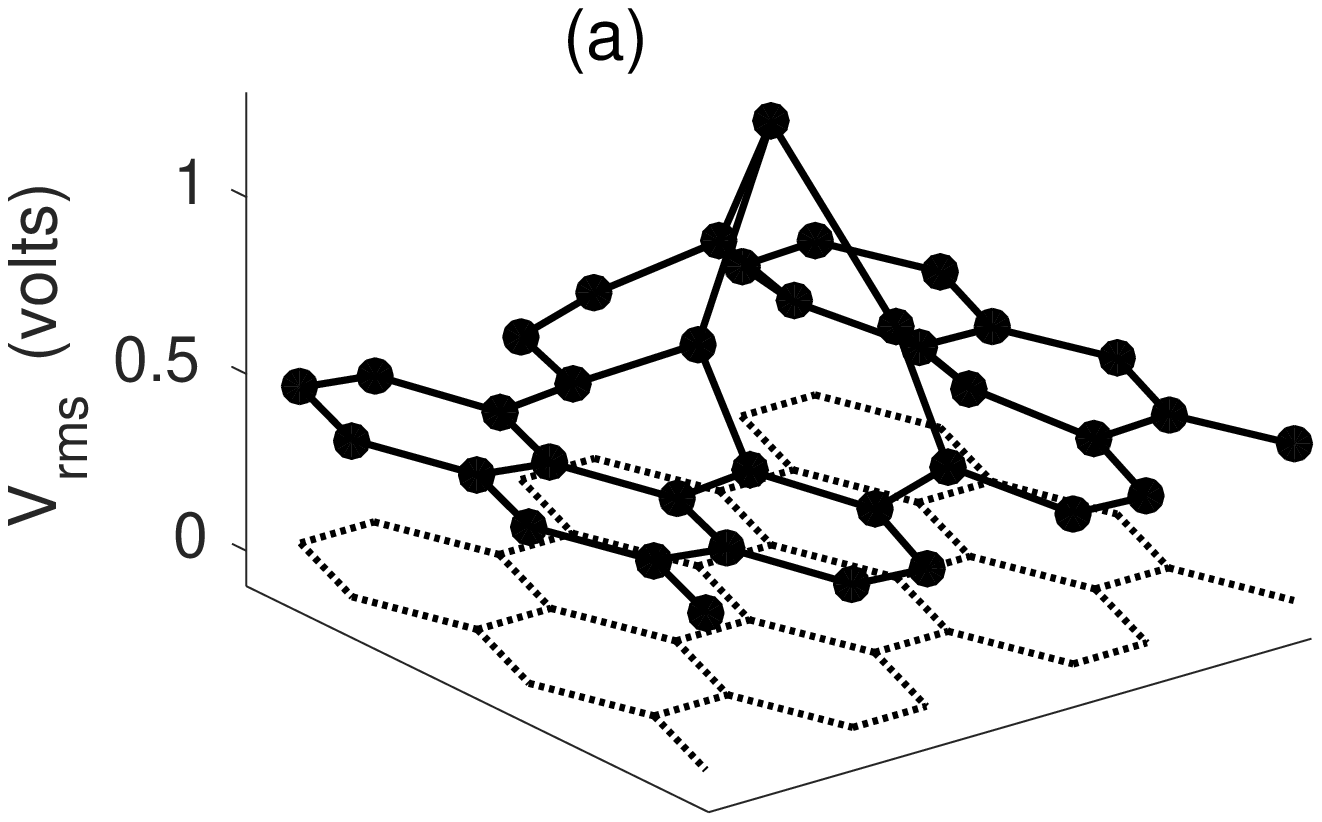}
\includegraphics[width=0.23\textwidth]{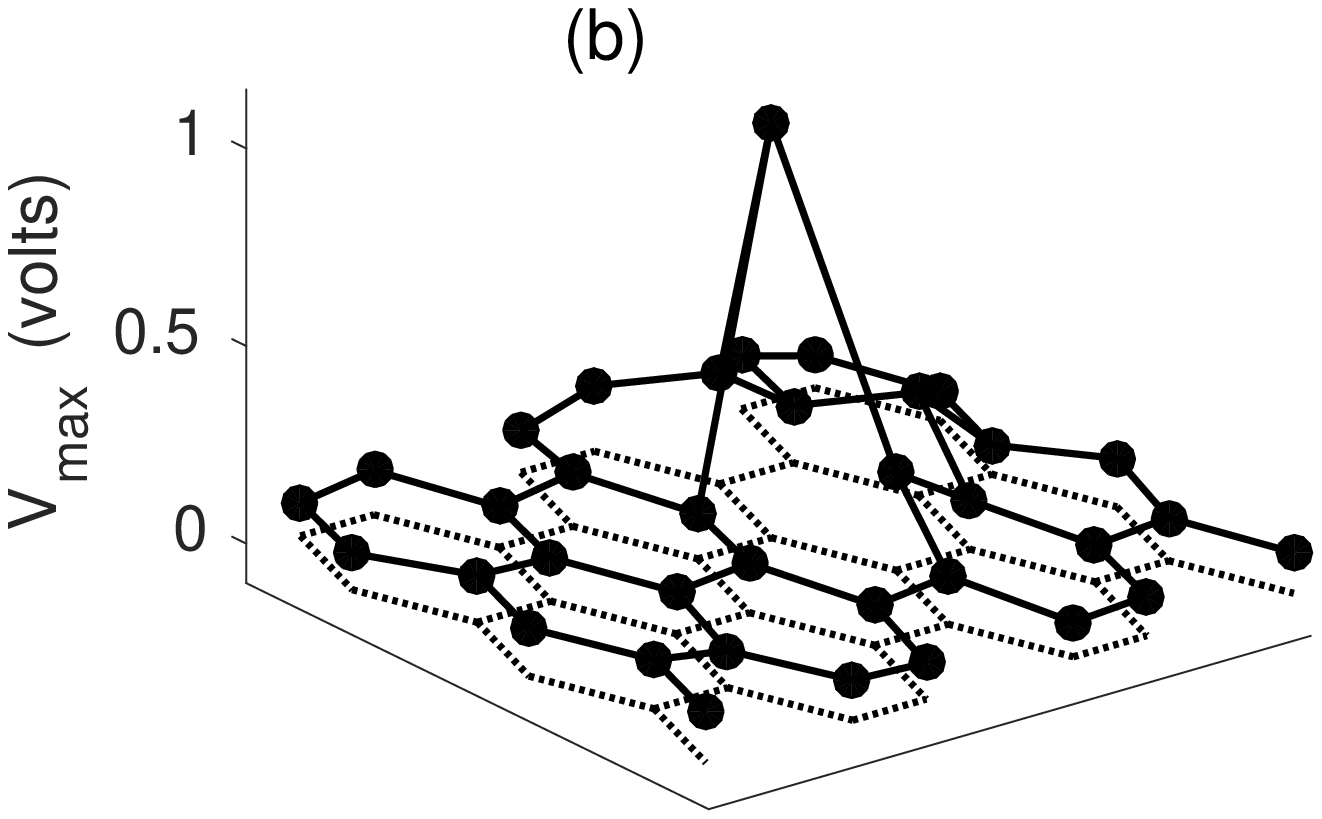}
\includegraphics[width=0.23\textwidth]{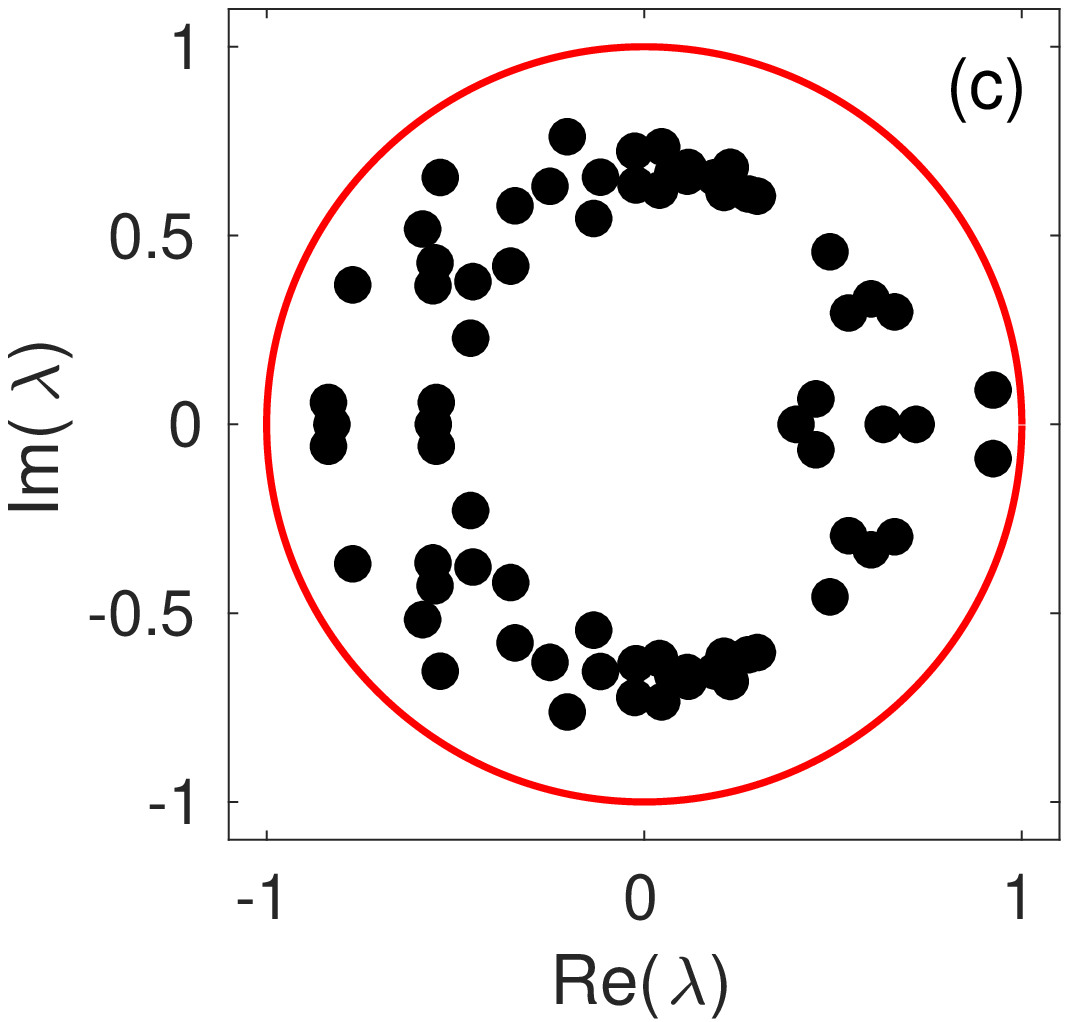}
\includegraphics[width=0.23\textwidth]{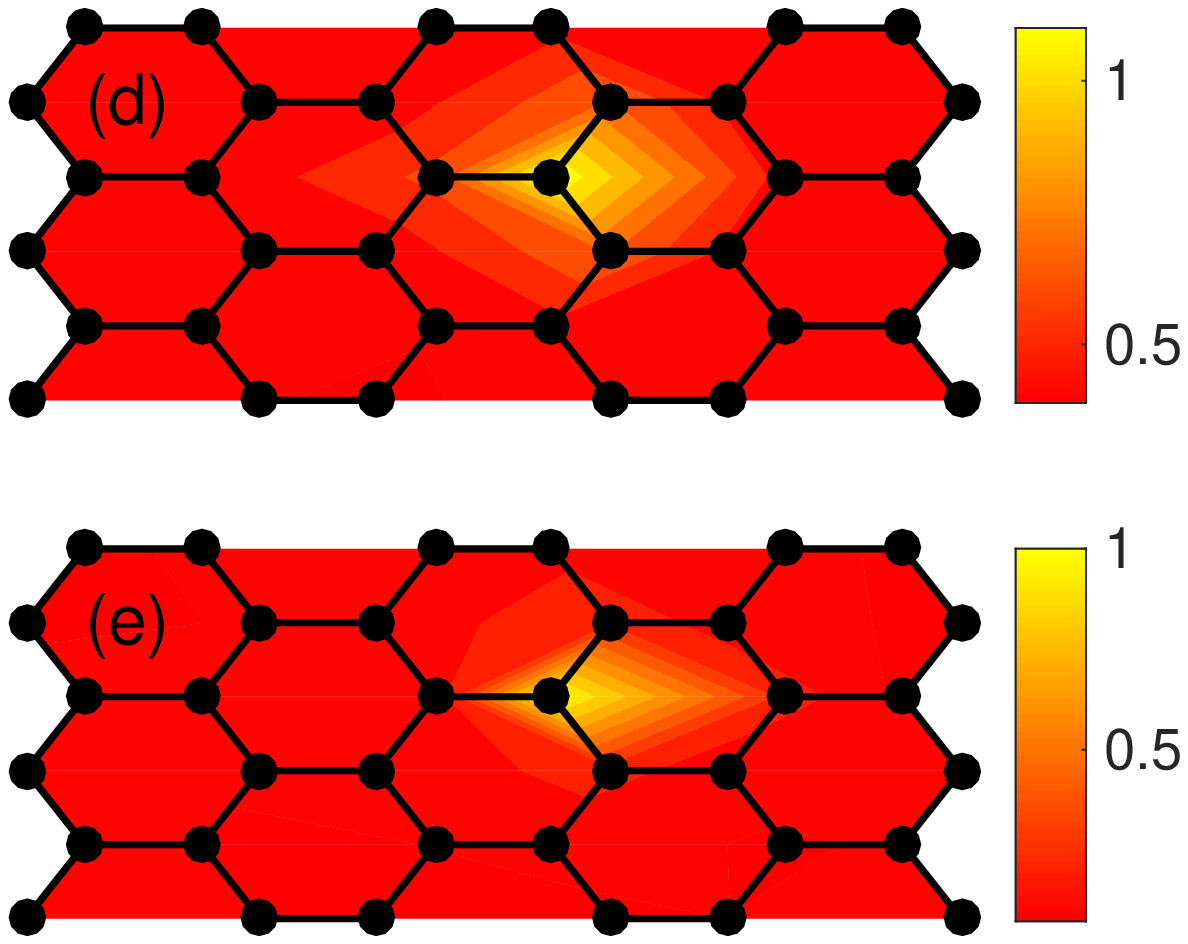}
\caption{(a) Numerical bulk breather (BB) profile in a $6\times 6$ lattice with periodic boundaries. (b) Experimental BB in the $6\times 6$ lattice with free boundary conditions. (c) Floquet multiplier spectrum corresponding to the numerical breather showing all the multipliers lying within the unit circle (and thus leading to the conclusion of spectral stability of such breathers). (d,e) Density plots corresponding to (a,b). $V_\mathrm{rms}$ in panels (a) and (b)stands for the root mean square of the voltage during a period. In both cases the {sinusoidal} driver amplitude was set to 2.1 V and the frequency was 278 kHz.}
\label{direct1}
\end{figure}

Having constructed the $6\times6$ honeycomb lattice of Fig.~\ref{lattice}, {the simplest experiment we can perform is to drive the lattice with a sinusoidal-wave profile and a frequency close the bottom of the linear
  modes band, as the lowest frequency mode is uniform ($\mathbf{k}=0$,
  i.e., the same wavevector as that of the driver)}. This is performed
in a progression of frequencies starting from outside (under) the
linear mode band and systematically increasing the frequency of the
drive. When the driver frequency is near the bottom of the linear
band (i.e. $f\lesssim f_\mathrm{min}$), we can generate
experimentally both BBs and EBs, where the latter seems to be the most
robust state between the two. Under the same conditions, in our theoretical model we find that only EBs exist. Alternatively, the use of periodic boundary conditions enables the existence of BBs for such frequencies. Fig.~\ref{direct1} shows a numerical bulk breather corresponding to a $6\times 6$ lattice with periodic boundaries and its Floquet multipliers spectrum (see e.g.~\cite{Aubry} for more details on Floquet analysis for discrete breathers). Recall that the existence of the corresponding multipliers solely within the unit circle for our driven/damped system indicates its spectral stability. In that figure, we also show the experimental BB obtained in the finite size lattice with free boundary conditions. In both cases the driver amplitude was set to 2.1 V and the frequency was 278 kHz. This is a representative example of such BBs within their relevant interval of existence (see also the discussion below).

\begin{figure}
\includegraphics[width=0.23\textwidth]{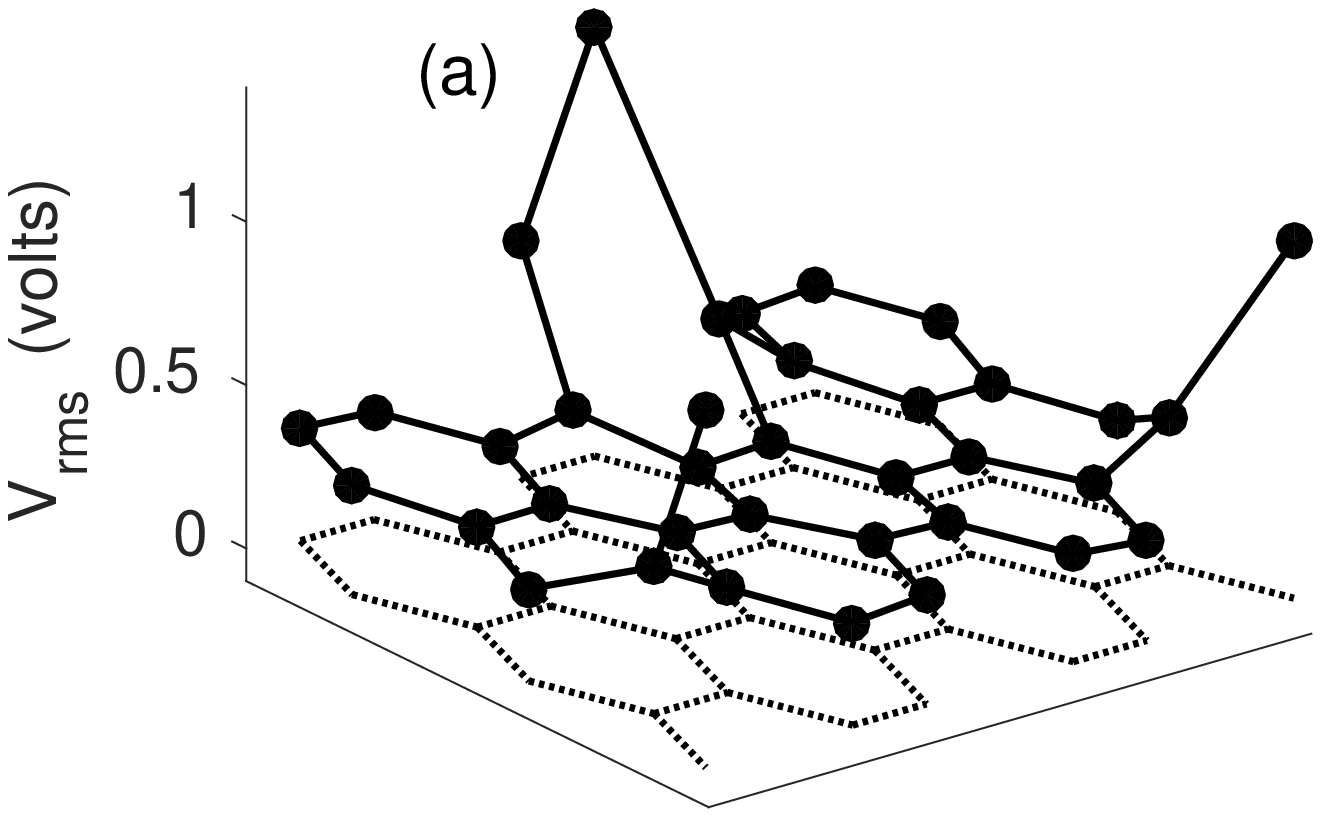}
\includegraphics[width=0.23\textwidth]{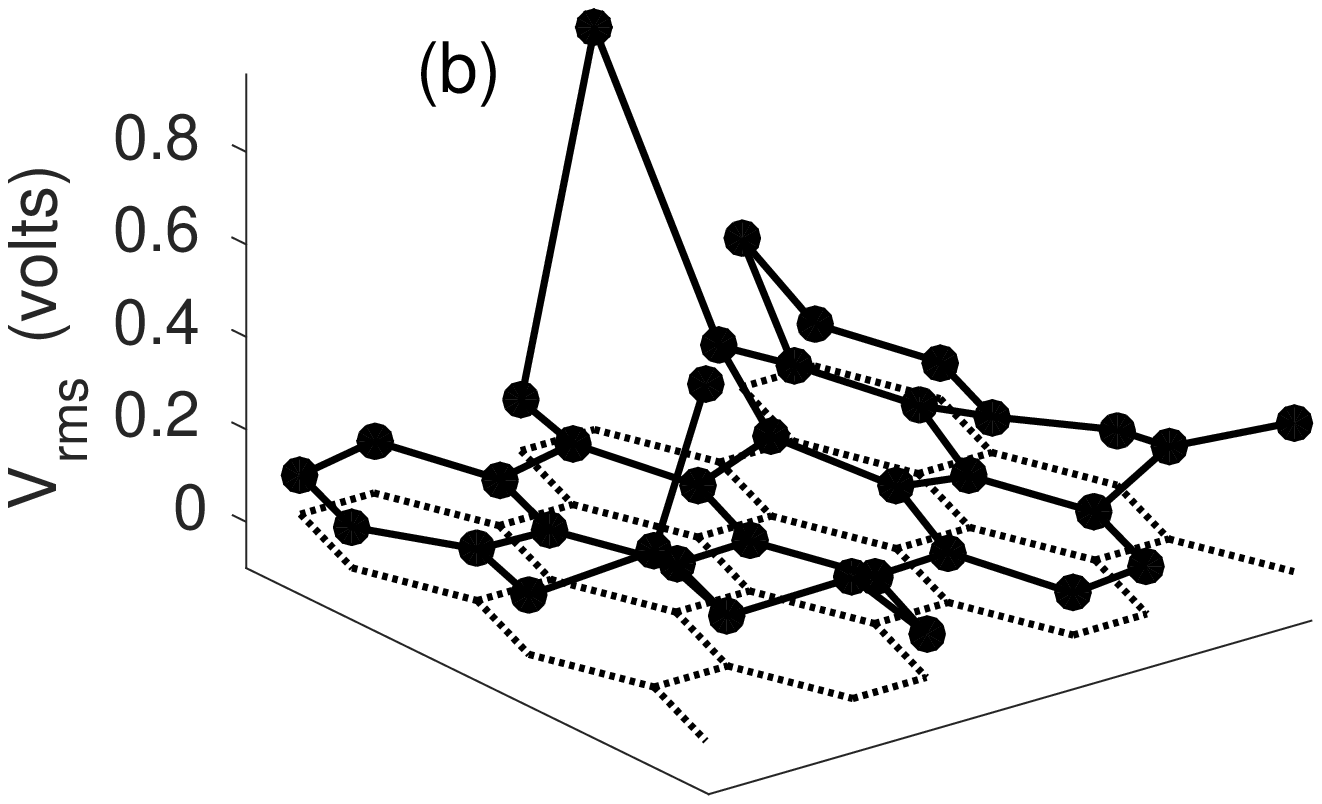}
\includegraphics[width=0.23\textwidth]{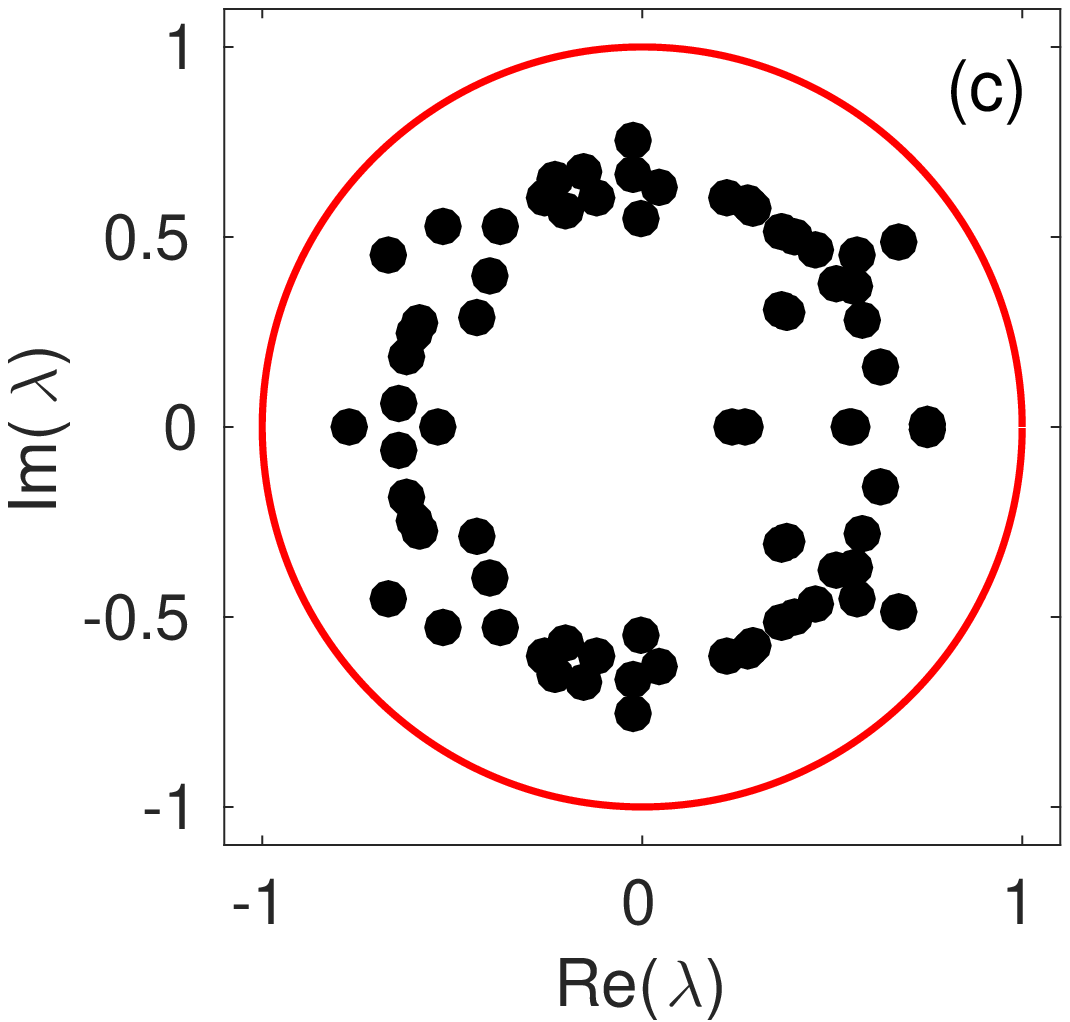}
\includegraphics[width=0.23\textwidth]{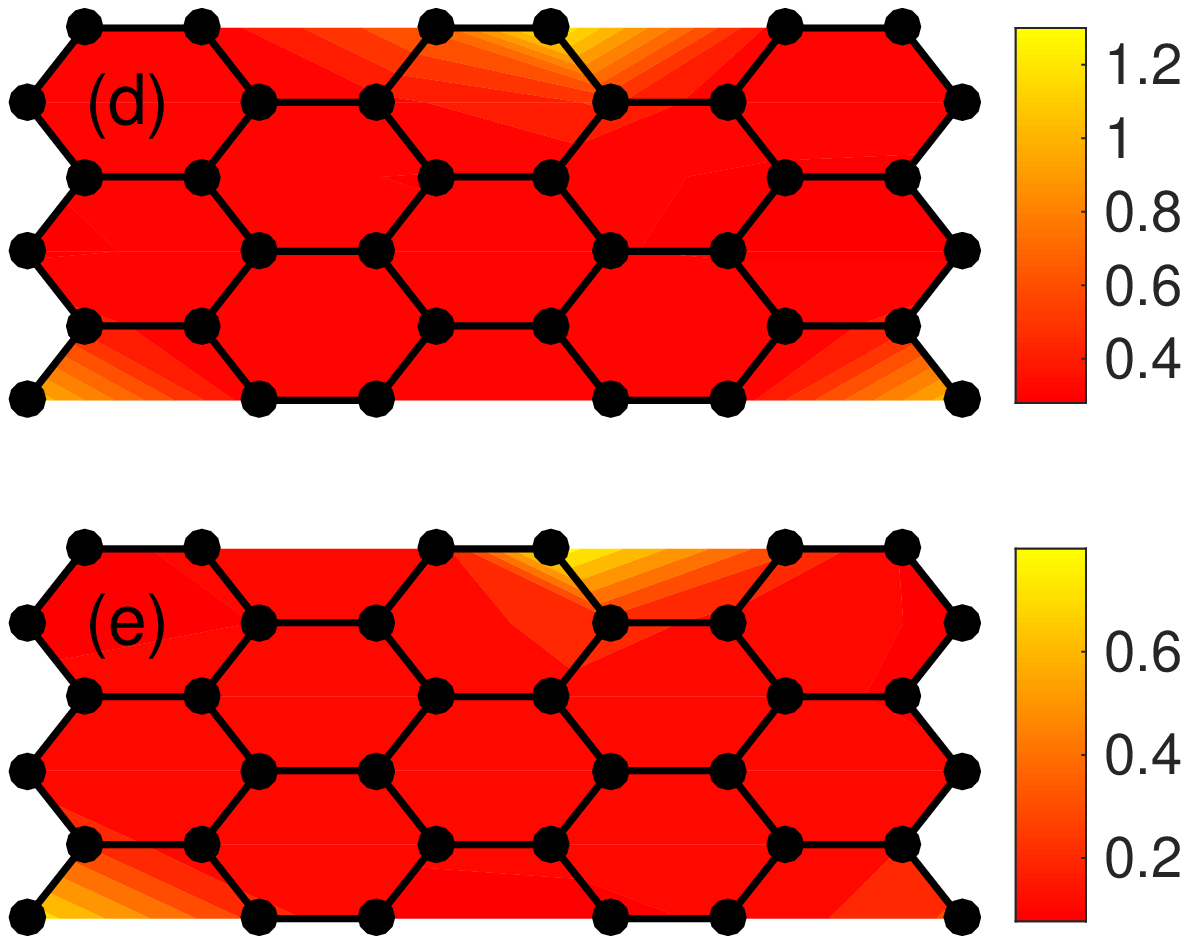}
\caption{Same as figure {\ref{direct1}} but for an edge breather (EB) in a free-boundary lattice.}
\label{direct2}
\end{figure}

Similarly, we can induce EBs which are, as indicated above, more
robust than BBs. Fig.~\ref{direct2} shows an example of the
theoretical and experimental features of an EB whose driving
parameters are the same as for the BB of Fig.~\ref{direct1}. The
existence of both kinds of solutions for the same system parameters
indicates the  multistability
of the system, given the different branches (bulk vs. edge) of
solutions. That is, the regions of existence for the different kinds
of breathers substantially overlap. The EBs are found to be somewhat
more stable in the following sense: as we lower either the frequency
or the amplitude of the driver (starting from 278 kHz and 2.1 V), the BB
will disappear first, before the EB ceases to exist. This means that
there is a small window in driving parameters where only edge
breathers can be stabilized. This finding, i.e. the wider range of
stabilization of the EB relative to the BB, has been also
experimentally observed in a chain of coupled pendula~\cite{pend}.

It should be mentioned that breathers can also be generated via
subharmonic driving. In that case, breathers (which are also denoted
as {\em subharmonic breathers}) are characterized by a core (i.e. the
peak and large amplitude nodes around it) oscillating with half of the
driver frequency whereas tails oscillate with the driving frequency
(see \cite{lars2}). Unlike what is observed in the experiments
(featuring both BBs and EBs), it seems that numerically only subharmonic
EBs are stable for this $6\times 6$ lattice with free boundaries (the
analysis of subharmonic breathers in larger lattices will be the
subject
of further studies). The observation of long-lived subharmonic BBs in the experiment may be due to small spatial inhomogeneities in the lattice facilitating their stabilization~\cite{electric2}. In terms of the subharmonic EBs, a good agreement is found regarding both their existence and their dynamical robustness.

\subsection{Driving near the Dirac point}

As the driver frequency is increased from around 280 kHz, we first
start producing more breathers in the lattice, as expected from
previous studies \cite{lars1}. Then, at higher frequencies, the
lattice response gradually weakens. The modes at the Dirac point (with
frequency 489.11 kHz) cannot be stimulated directly; this is a
consequence of the fact that the Dirac mode possesses a non-zero
wavevector, whereas the driver is associated with the zero wavevector.

At higher driving amplitudes, however, lattice response in the
vicinity of the Dirac point can be induced via subharmonic
driving. Both subharmonic BBs and EBs, which are excited close to the
Dirac point frequency, are similar to the ones produced via direct
driving although the oscillation frequency {of the excited sites}
is the half of that of the driver {and the breather tail} (see also
Ref. \cite{lars2}). If the (sinusoidal) driver amplitude is increased to 9 V, a
clear lattice subharmonic response is observed in the range
$(530,695)$ kHz. When using a square-wave driving profile, this range
is slightly expanded; this phenomenon, which has been reported
recently for non-sinusoidal drivings, is related to the enhancement of
the ``mechanical'' impulse transmitted to the lattice from the driver,
facilitating the generation of stationary breathers in experiments,
as well as in numerical computations~\cite{impulse1}.

\begin{figure}
\includegraphics[width=0.23\textwidth]{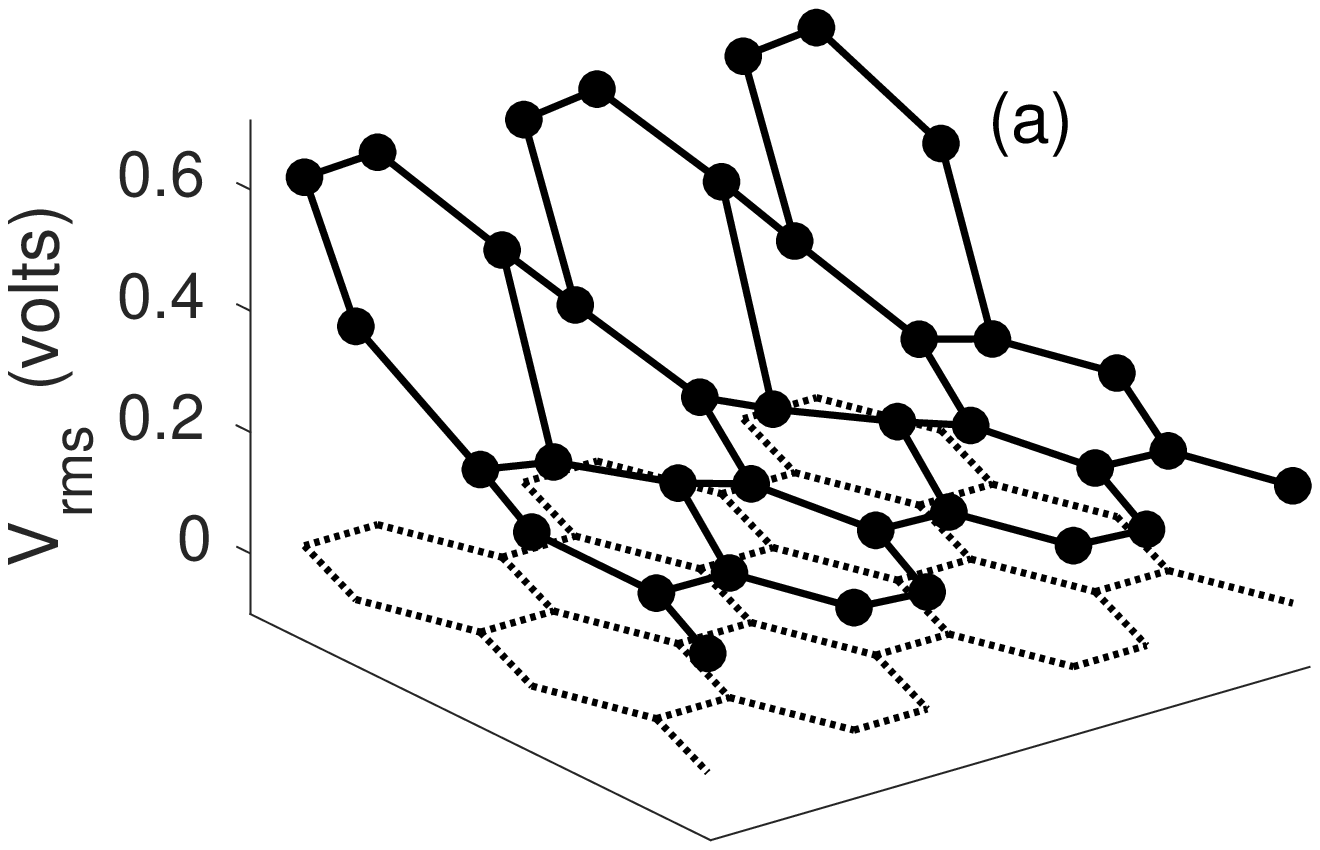}
\includegraphics[width=0.23\textwidth]{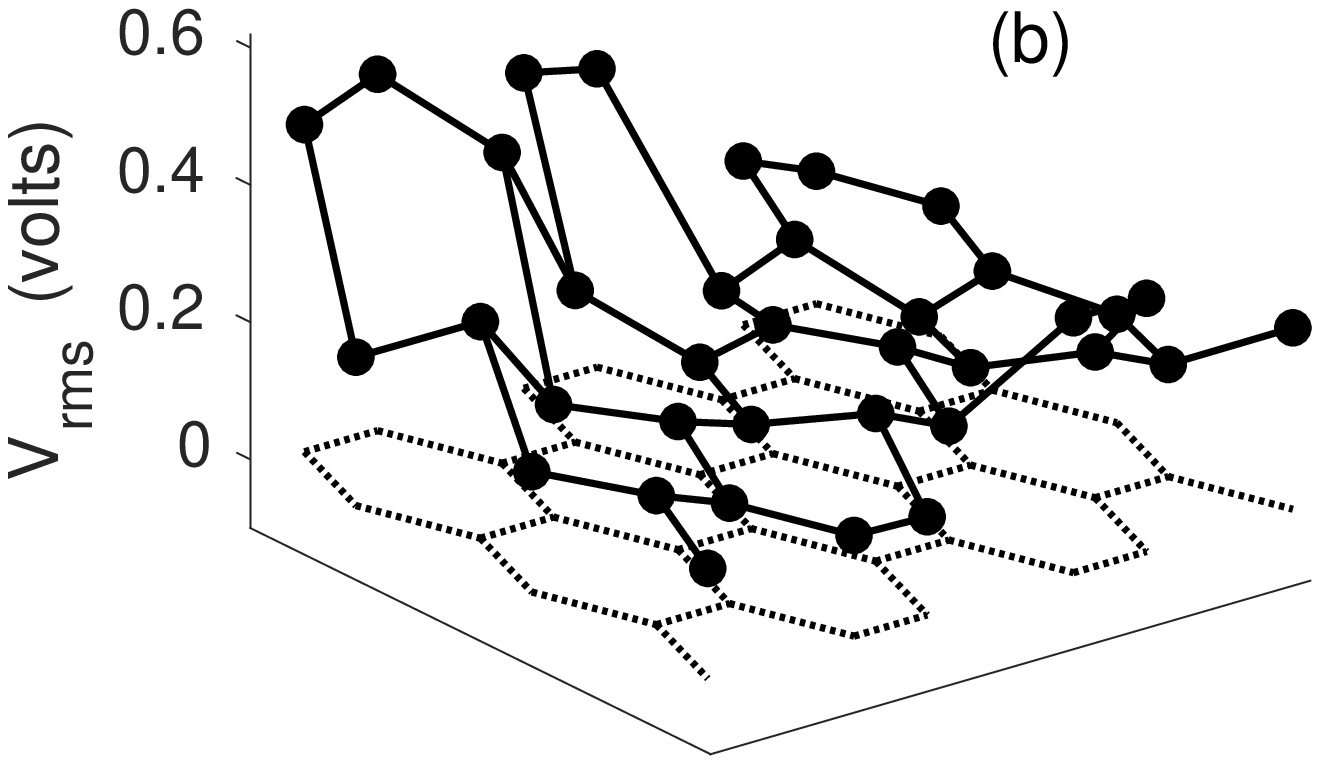}
\includegraphics[width=0.23\textwidth]{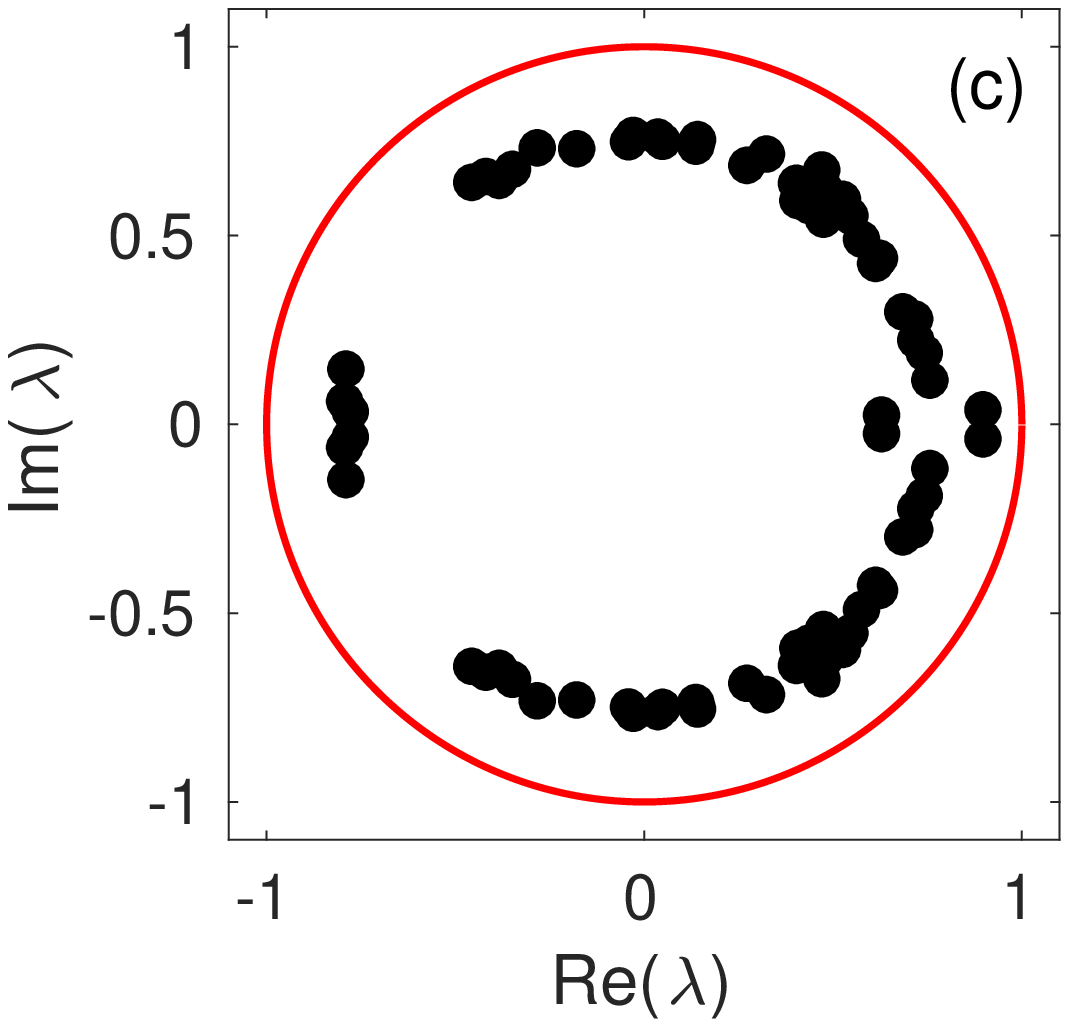}
\includegraphics[width=0.23\textwidth]{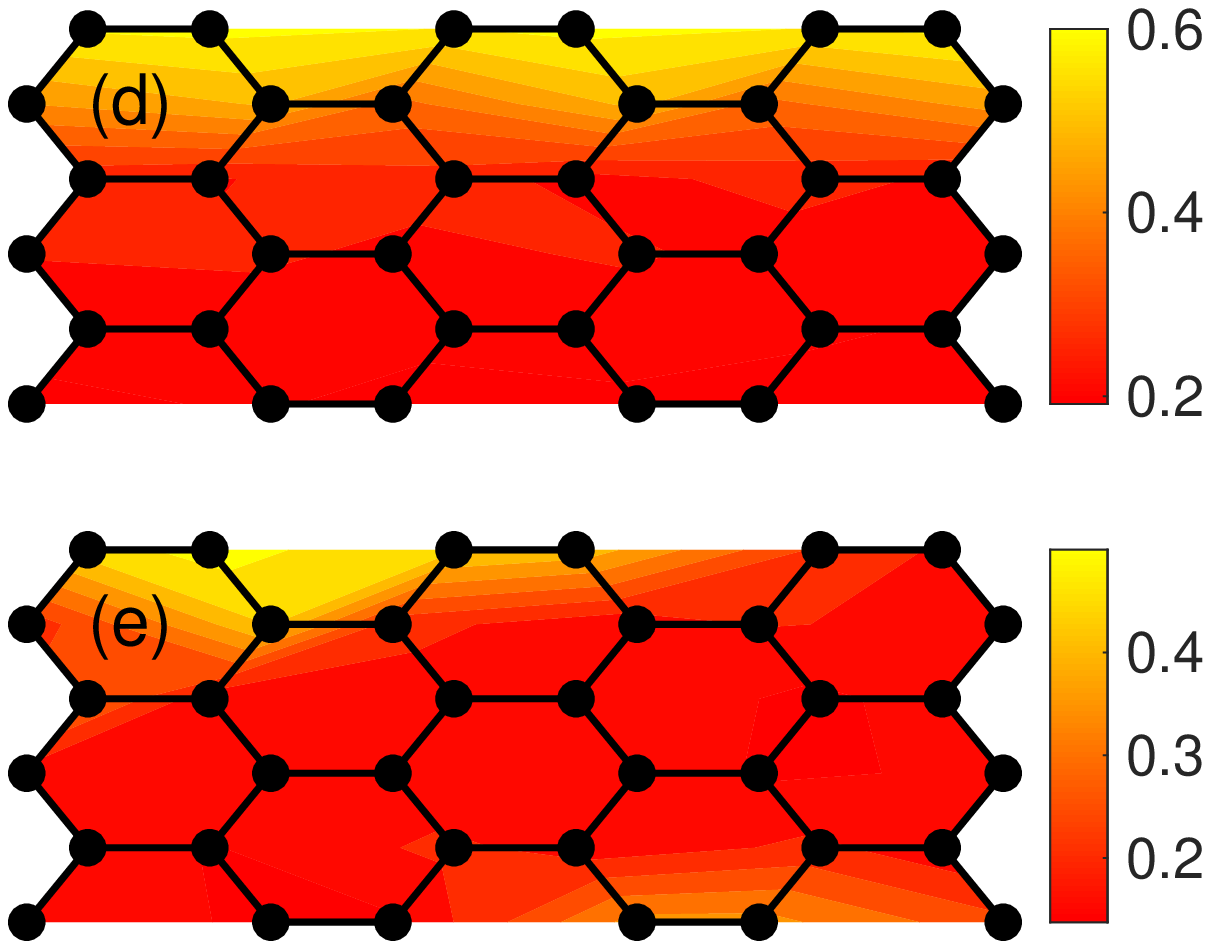}
\caption{Same as Fig.~\ref{direct2} but for a subharmonic armchair-EB with a { square-wave driving profile of amplitude 9 V.} The driving frequency was 960 kHz in numerics and 940 kHz in experiments}
\label{dirac1}
\end{figure}

Above the upper edge of this frequency window, the lattice response goes to zero again (at least for the uniform driving used experimentally). Then, starting at 886 kHz and using a square-wave driving at 9 V (no subharmonic response has been found for this range of frequencies using sinusoidal driving), an EB appears firstly along the armchair edge of the honeycomb lattice (also dubbed as armchair-EB). In experiments this mode persists up to a frequency of 940 kHz, corresponding to a response frequency at the breather peak of 470 kHz. Numerical simulations show similar results, as represented in Figure~\ref{dirac1}.

\begin{figure}
\includegraphics[width=0.23\textwidth]{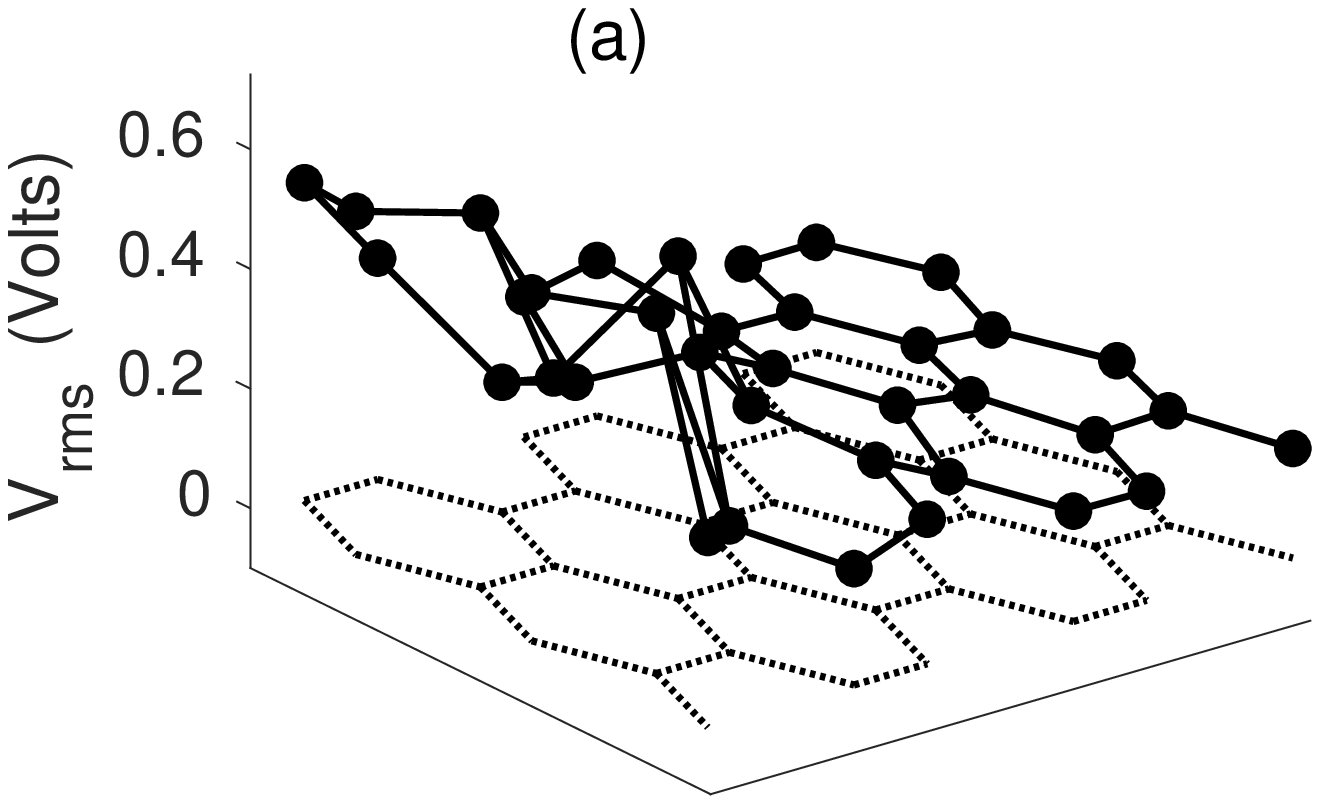}
\includegraphics[width=0.23\textwidth]{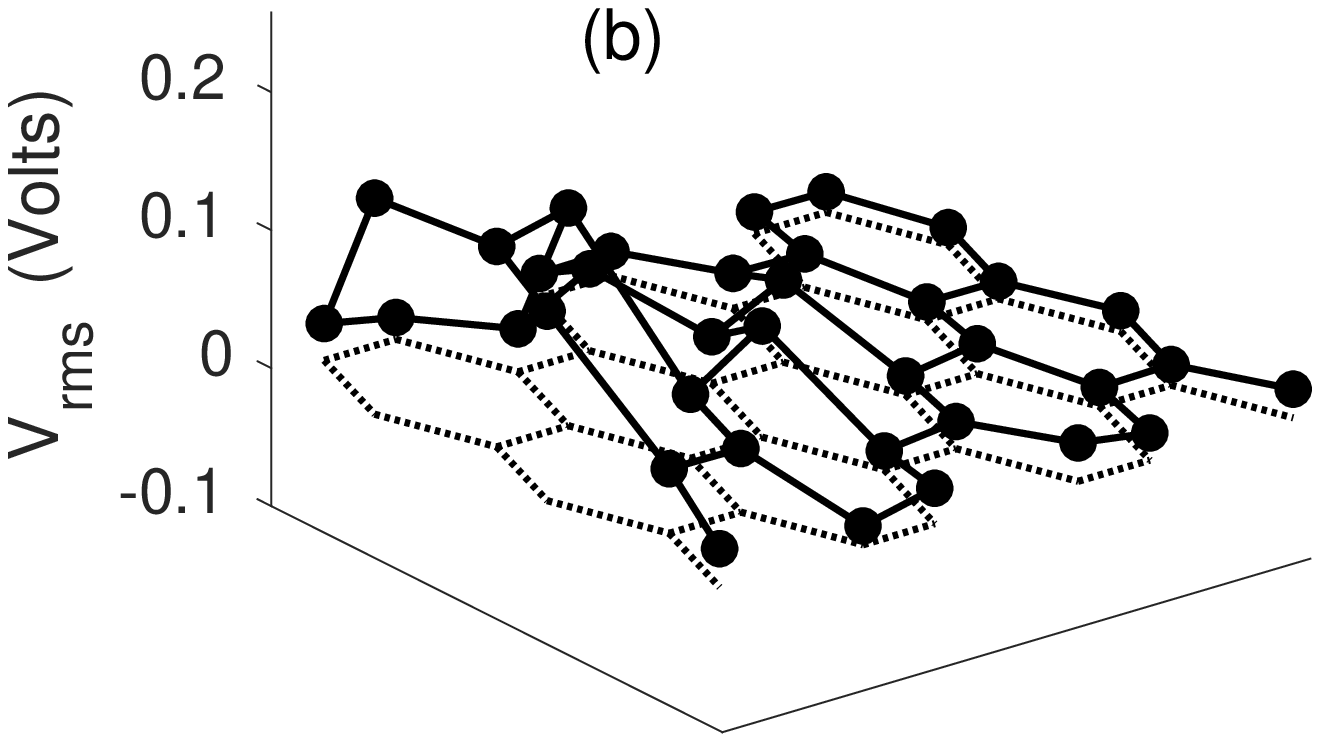}
\includegraphics[width=0.23\textwidth]{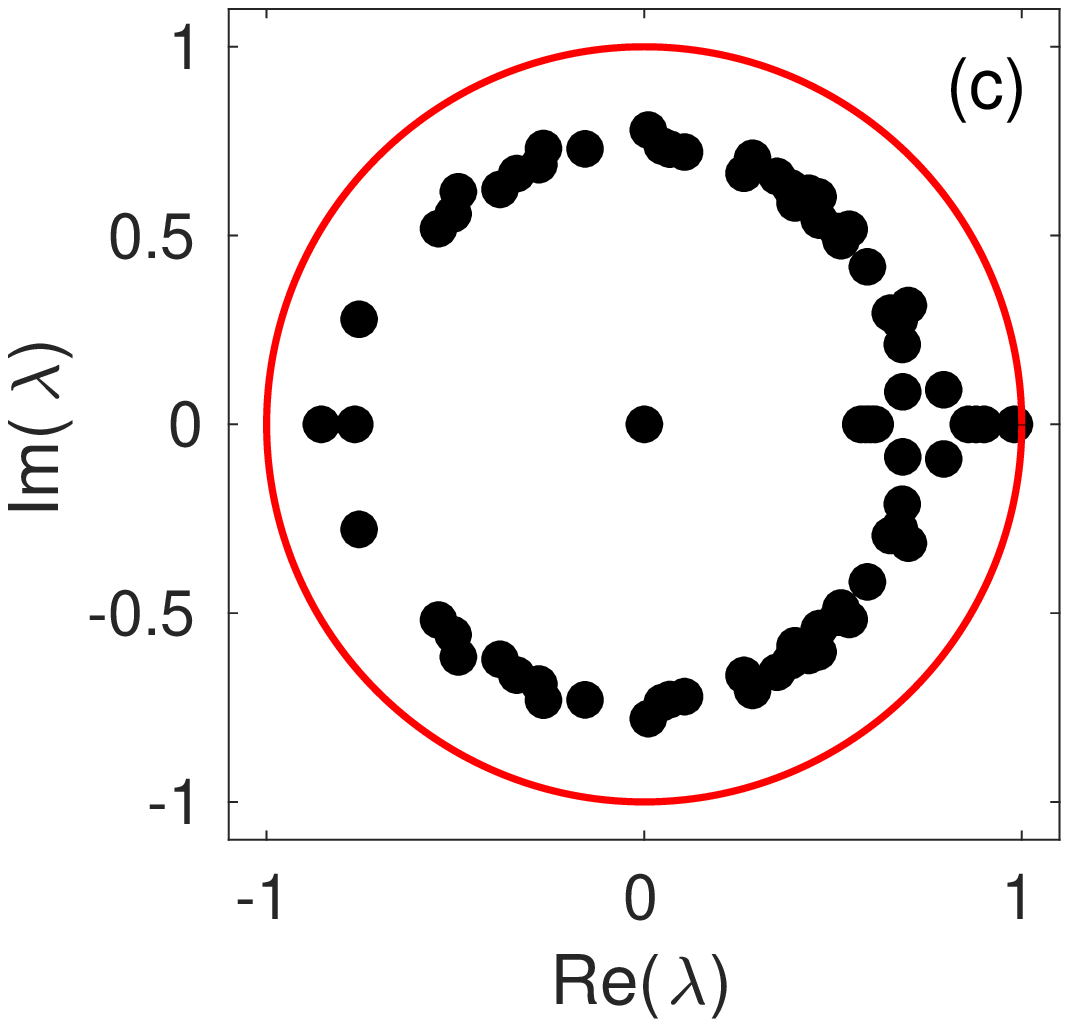}
\includegraphics[width=0.23\textwidth]{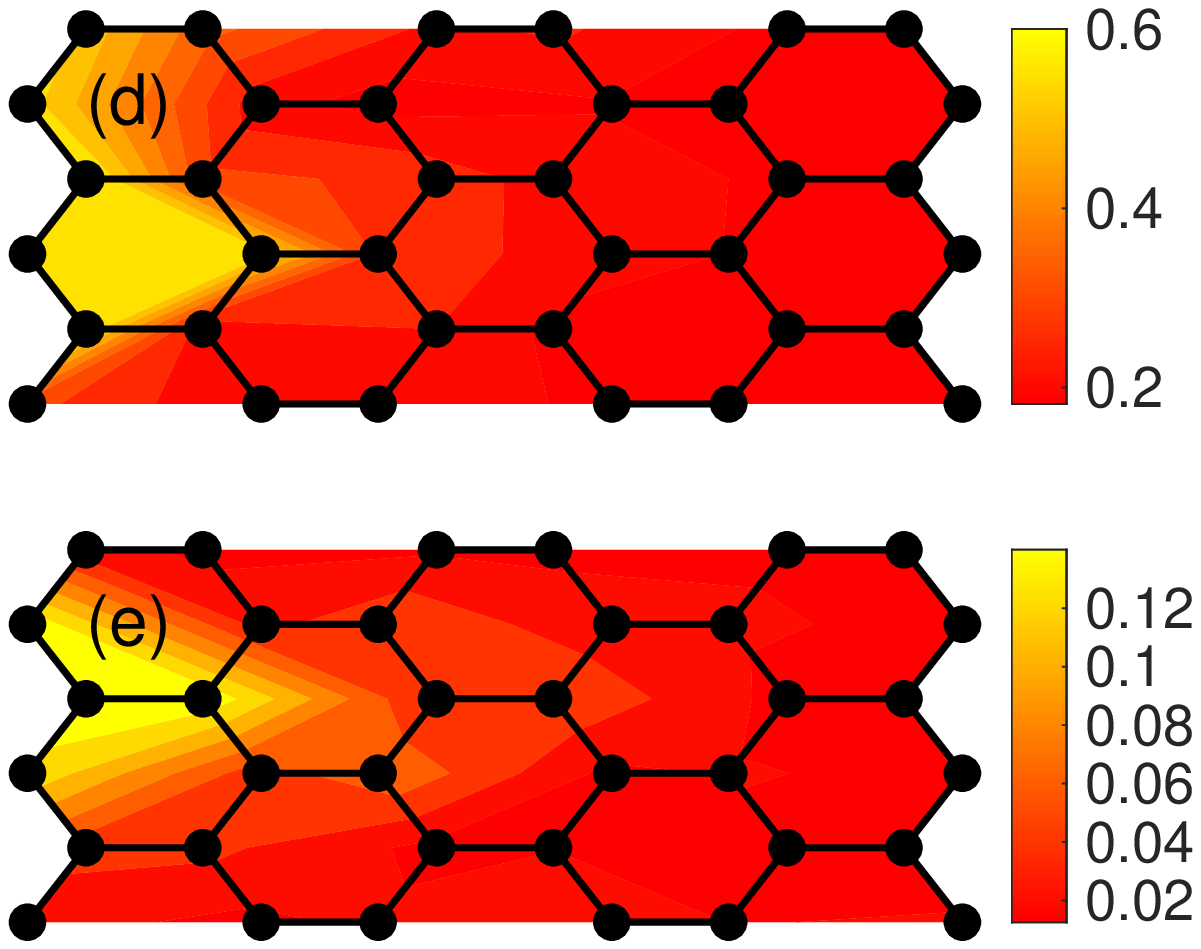}
\caption{Same as Fig.~\ref{dirac1} but for a subharmonic zig-zag-EB. The driving frequency was 1000 kHz in numerics and 960 kHz in experiments}
\label{dirac2}
\end{figure}

At a driver frequency of 950 kHz, we witness an abrupt switch to an EB
along the zig-zag edge of the lattice (also dubbed as
zig-zag-EB). Such a breather, whose main features are shown in
Fig.~\ref{dirac2}, persists up to a driving frequency of 967 kHz. In
general, numerics are in qualitative agreement with the experiments. The switch
between these two types of edge breathers as the frequency is
adiabatically increased is very reproducible. This clearly suggests
the different intervals of stability of the two edge configurations,
indicating which one is the system's lower energy state for the different frequency regimes.

\begin{figure}[h]
\includegraphics[width=0.23\textwidth]{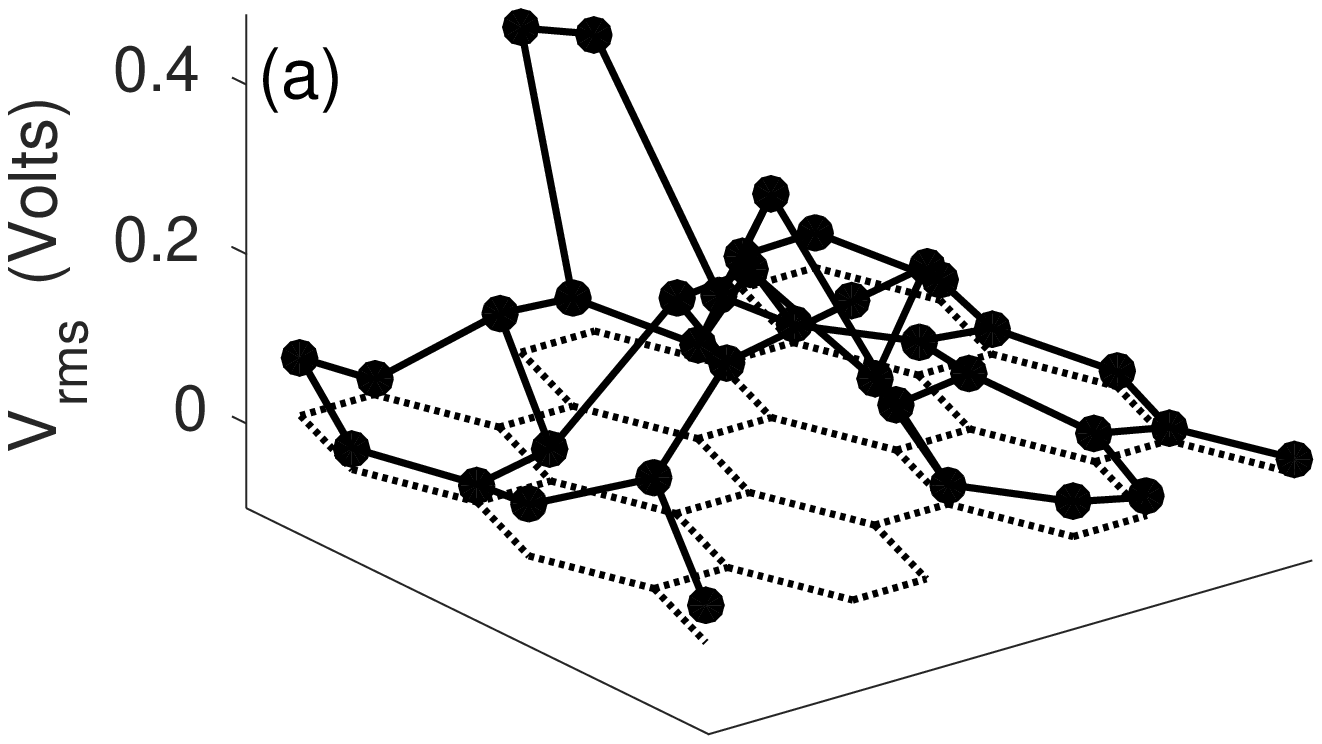}
\includegraphics[width=0.23\textwidth]{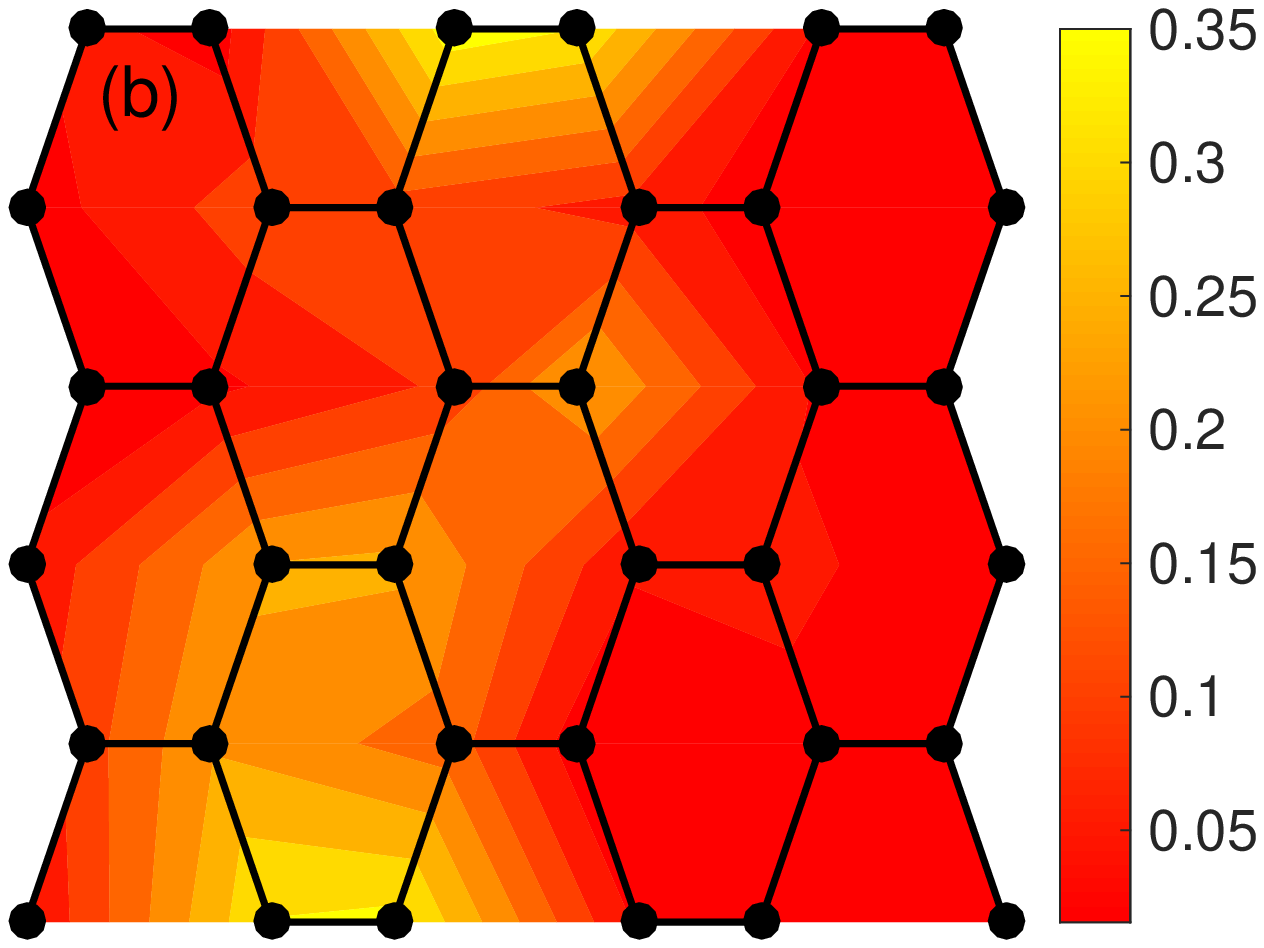}
\includegraphics[width=0.23\textwidth]{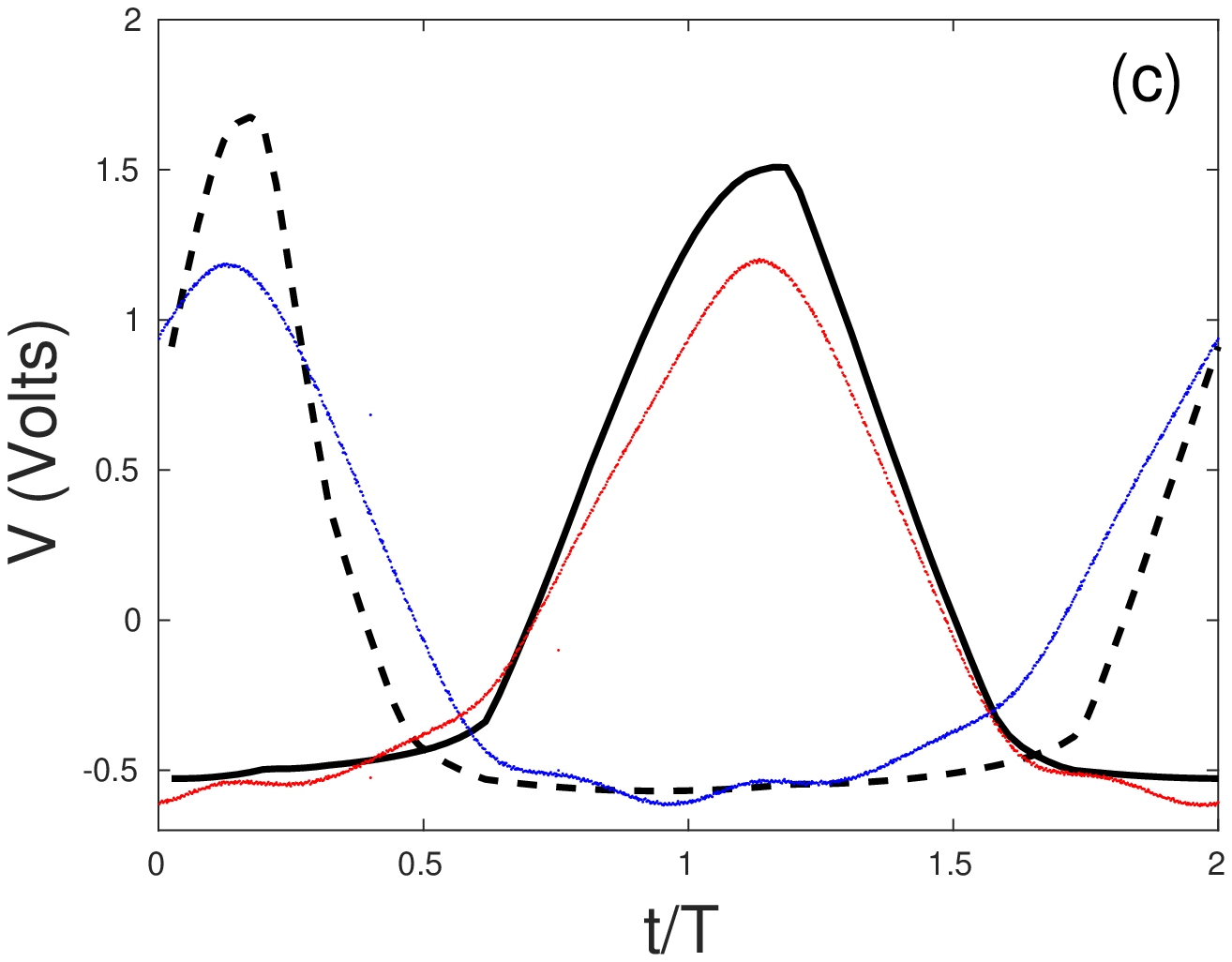}
\caption{(a) Experimental subharmonic armchair-EB profile in a $6\times 6$ lattice with free boundaries and (b) its corresponding density plot. The	{square-wave}	driving amplitude was set to 9 V and the frequency to 886 kHz. (c) Time dependence of the voltage at the two largest amplitude nodes on the top of the lattice: numerical simulations are shown by black continuous and dashed lines, whereas experimental data are shown in blue and red lines.}
\label{dirac3}
\end{figure}

Figure \ref{dirac3} examines the armchair-EB more closely. Here we showcase the most nonlinear, and therefore most localized version of that mode at a driving frequency of 886 kHz. As it can be seen from the experimental pattern depicted in panel (b), there are two of such EBs at the opposite sides of the lattice. Comparing to Fig.~\ref{lattice}, one can observe in both cases a sharp localization of the energy at the outermost node-pairs along the armchair edges. It is evident, from the detailed time-dependent oscillations pattern of panel (c), that the two largest-amplitude nodes in the top of the lattice oscillate in anti-phase, indicating that $\mathbf{k}\neq\mathbf{0}$ for this EB. Once again, numerical simulations are in good agreement with experimental results. Similar features (not shown here) are shared with zig-zag-EBs.

\begin{figure}[h]
\includegraphics[width=0.45\textwidth]{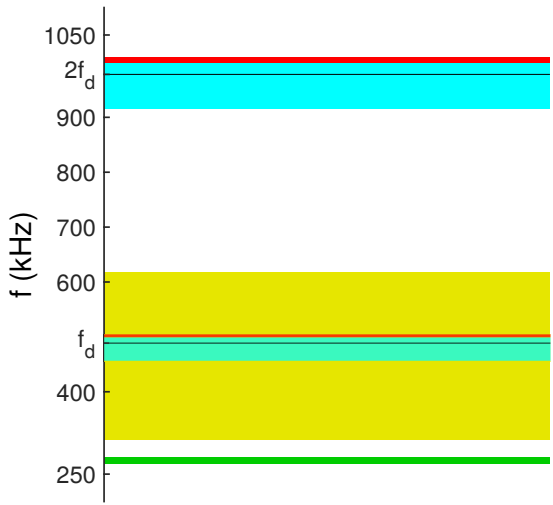}
\caption{Stability range versus driving frequency for different kinds
  of numerically calculated EBs together with the linear modes band,
  which is represented as a yellow region; Dirac point ($f_d$) and
  twice its value are depicted as black horizontal line. Green region
  corresponds to (direct-driven) EBs with $V_d=2.1$ V
  (cfr. Fig.~\ref{direct2}). (b) Subharmonic armchair-EBs driven by
  $V_d=9$ V (cfr. Fig.~\ref{dirac1}) is represented by the blue
  region, and the lighter blue region overlapping part of the linear
  modes band corresponds to the subharmonic oscillations of the
  breather peak. Similarly to the armchair-EB case, the red region
  and the lighter red one in the linear modes band correspond to subharmonic zig-zag-EBs.}
\label{cont}
\end{figure}

\begin{figure}
\includegraphics[width=0.45\textwidth]{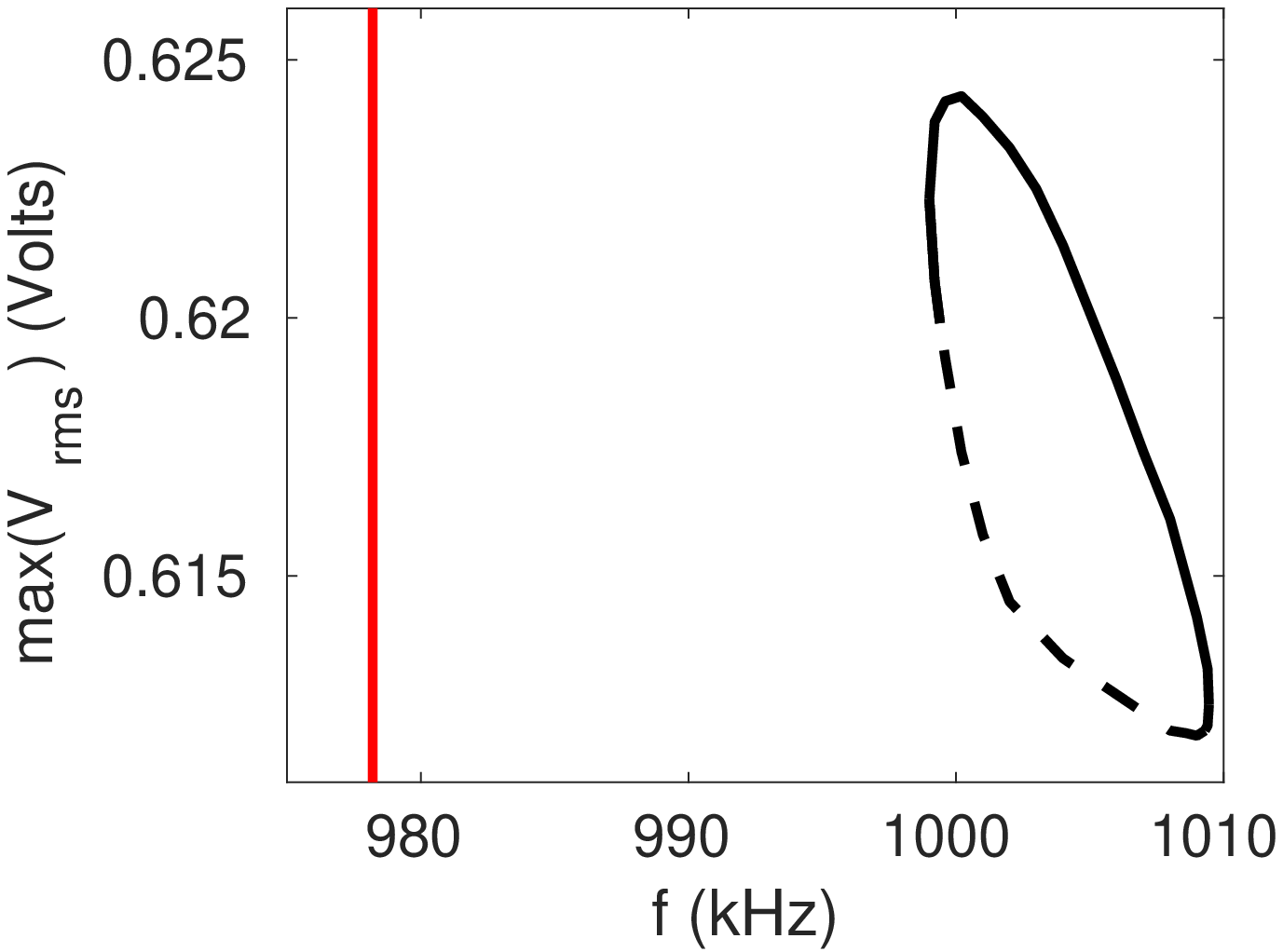}
\caption{$V_\mathrm{rms}$ on peak node of the numerically calculated
  subharmonic EBs (solutions in red and blue at Fig.~\ref{cont}), as a
  function of the frequency. Stable solutions are depicted as
  continuous lines and unstable solutions as dotted lines. The red line marks twice Dirac frequency ($2f_d$).}
\label{cont2}
\end{figure}

Edge breathers around the above mentioned frequency ranges
{i.e. $f\in[886-967]$ kHz} were not found to occur in the square
lattice, thus seeming to be particular to the honeycomb geometry and
its associated boundary geometry. Furthermore, it is an interesting
fact that zig-zag-EBs and armchair-EBs are found to occupy such
distinct frequency bands around the calculated Dirac frequency, as
shown in Fig.~\ref{cont}. We do not get any BB with this set of
boundary
conditions;
rather, only EBs arise. An example of a bifurcation diagram featuring a pair of saddle-node bifurcations for numerically calculated subharmonic EBs is shown in Fig.~\ref{cont2}. In general, we have found that this scenario of saddle-node bifurcations for the destabilization of solutions is fairly generic (results not shown here). Analyzing such features in more detail turns out to be a rather delicate task due to the existence of numerous branches in a complex bifurcation diagram. For this reason, a more exhaustive study is deferred to a future publication.

\section{Conclusions \& Future Work}

Naturally, the above findings constitute only a first step in the emerging rich study of localized modes and the breathing dynamics in honeycomb electrical lattices. Here, we have explored the arguably most canonical and experimentally more straightforwardly tractable case of a uniform drive of zero wavevector. At frequencies below the linear band, we have found that this drive leads to the formation of bulk, as well as edge breathers, with the latter being more robust than the former. However, our most significant finding concerns to the subharmonic drive in the vicinity of twice the frequency of the Dirac point. There, depending on the frequency interval, both armchair and zigzag edge breathers can arise, with each one appearing as the stable state in a respective interval frame.

There are numerous questions that still remain worthwhile to answer. Is it possible to achieve more elaborate forms of driving so as to excite higher
order states? At the same time, it appears to be relevant to develop a systematic continuation analysis, e.g., at the Hamiltonian level of the corresponding linear and also subharmonic waves and their expected role in the bifurcation diagram. Understanding whether these edge states enjoy topologically induced
propagation properties (e.g. through the lattice boundary) is an important question worth considering in its own right. Furthermore, the experimental
tractability of electrical lattices renders them interesting candidates for formulating additional lattices with intriguing topological properties
such as, e.g., some of the artificial flat band systems recently summarized in~\cite{flach2}.

\vspace{2mm}

{\it Acknowledgements.}  This
material is based upon work supported by the US National Science
Foundation under Grants No. PHY-1602994 and DMS-1809074
(PGK). PGK also acknowledges support from the Leverhulme Trust via a
Visiting Fellowship and the Mathematical Institute of the University
of Oxford for its hospitality during part of this work. {J.C.-M. was supported by MAT2016-
79866-R project (AEI/FEDER, UE). {FP visited Dickinson College with support from
	the VI Plan Propio of the University of Seville (VI PPITUS). FP also acknowledges Dickinson for its hospitality.} }

\end{document}